\documentclass[12pt,a4paper]{article}

  \usepackage{a4wide}
  \usepackage{latexsym}
  \usepackage{epsf}
  \usepackage{amssymb}
  \usepackage{graphicx}
  \usepackage{amsmath, cite}
  \usepackage{slashed,epsfig}
  \include{braket}
  \usepackage{bbm}

\begin{document}

\begin{flushright}
\small
UG-06-02\\
\date \\
\normalsize
\end{flushright}

\begin{center}

\vspace{.7cm}

{\LARGE {\bf Dynamics of Generalized Assisted Inflation}} \\

\vspace{1.2cm}

\begin{center}
  Jelle Hartong, Andr\'e Ploegh, Thomas Van Riet and Dennis B. Westra  \\[3mm]
   {\small\slshape
   Centre for Theoretical Physics, University of Groningen, Nijenborgh
  4,\\
9747 AG Groningen, The Netherlands \\
  {\upshape\ttfamily j.hartong, a.r.ploegh, t.van.riet, d.b.westra@rug.nl}}\\[3mm]

\end{center}

\vspace{5mm}

{\bf Abstract}

\end{center}

\begin{quotation}

\small We study the dynamics of multiple scalar fields and a
barotropic fluid in an FLRW-universe. The scalar potential is a
sum of exponentials. All critical points are constructed and these
include scaling and de Sitter solutions. A stability analysis of
the critical points is performed for generalized assisted
inflation, which is an extension of assisted inflation where the
fields mutually interact. Effects in generalized assisted
inflation which differ from assisted inflation are emphasized. One
such a difference is that an (inflationary) attractor can exist if
some of the exponential terms in the potential are negative.

\end{quotation}

\newpage

\pagestyle{plain}

\section{Introduction}

Scalar fields can violate the strong energy condition, which is a
necessary condition for inflation and present-day acceleration. An
alternative way to violate the strong energy condition is pure
vacuum energy, that is, a positive cosmological constant
$\Lambda$. Observations do not exclude either possibility, but
scalar fields provide more freedom to evade typical fine-tuning
problems posed by a cosmological constant such as the smallness of
$\Lambda$ or the cosmic coincidence problem. On the other hand it
is argued that scalar fields generate new problems not posed by a
cosmological constant \cite{Padmanabhan:2005ur}.

In this paper we investigate scalar field models and we confine
ourselves to models where the scalar fields have a canonical
kinetic term and the potential is a sum of exponential terms where
the exponent is a linear combination of the scalars. The reason
for this choice is twofold. First of all, these models allow to
find exact solutions, which correspond to critical points in an
autonomous system \footnote{For a review on dynamical systems in
cosmology see \cite{Coley:1999uh}.}. The second reason is that
exponential potentials arise in models motivated by string theory
such as supergravities obtained from dimensional reduction (see
for instance
\cite{Ohta:2004wk,Bergshoeff:2003vb,Karthauser:2006wb,Bremer:1998zp,Green:1999vv}
and references therein), descriptions of brane interactions
\cite{Mazumdar:2001mm,Becker:2005sg,Ward:2005ti}, nonperturbative
effects and the effective description of string gas cosmology (see
for instance \cite{Berndsen:2005qq}).

The simplest example is the single-field model with one positive
exponential potential, which was shown to have a stable powerlaw
inflationary solution when the potential is sufficiently flat
\cite{Halliwell:1986ja}. Later on this model was extended by
including a barotropic fluid, negative exponentials, spatial
curvature and combinations thereof
\cite{Wetterich:1987fm,Coley:1997nk,Copeland:1997et,Heard:2002dr,vandenHoogen:1999qq}.
In all cases it is possible to rewrite the equations of motion as
an autonomous system. Some critical points correspond to so-called
scaling solutions. Scaling solutions are defined as solutions
where different constituents of the universe are in coexistence
such that the ratios of their energy densities stays constant
during evolution.

Since supergravities typically contain many scalars we are
interested in the extension of these models to multiple scalars.
In reference \cite{Liddle:1998jc} the authors consider a potential
that is a sum of exponentials with each exponential containing a
different scalar. The main result is that if each separate
exponential is too steep to drive inflation the system can still
have an inflationary attractor provided the number of fields is
sufficiently large. This effect is known as assisted inflation. In
reference \cite{Coley:1999mj} this system is extended to include a
barotropic fluid and spatial curvature.

In assisted inflation there is no mutual interaction
(cross-coupling) between the scalars. To understand the effects of
mutually interacting scalars reference \cite{vandenHoogen:2000cf}
investigates a model with one exponential term containing multiple
scalars. But as we will argue this theory can be rewritten as a
theory without cross-coupling. Mutual interaction of the scalar
fields can only be present when there are more exponentials terms
in the potential.

As shown in \cite{Collinucci:2004iw} multiple exponential
potentials are naturally divided into two classes:
\begin{enumerate}
\item The minimal number of scalar fields which can occur in the
potential equals the number of exponentials. \item The minimal
number of scalar fields which can occur in the potential is less
than the number of exponentials.
\end{enumerate}
A more precise definition can be found in section \ref{system}.
The first class can have scaling solutions and the second class
can have scaling and de Sitter solutions.

Here we shall mostly be interested in class 1. This class is
termed generalized assisted inflation in \cite{Copeland:1999cs}.
Generalized assisted inflation includes assisted inflation as a
special case where there is no cross-coupling. In reference
\cite{Copeland:1999cs} some exact solutions of generalized
assisted inflation were given, which were later shown to
correspond to a subset of all possible critical points
\cite{Collinucci:2004iw}.

In contrast to assisted inflation, generalized assisted inflation
is poorly investigated. Therefore we extend the results of
\cite{Copeland:1999cs,Collinucci:2004iw} in two ways. First, by
constructing all possible critical points of the model containing
the most general multiple exponential potential with a barotropic
fluid and spatial curvature. The second and more important
extension is a stability analysis of the critical points in
generalized assisted inflation.

The paper is organized as follows. In section 2 the equations of
motion are given in a flat FLRW-background and are then rewritten
as an autonomous system. In section 3 we construct all critical
points for the most general multiple exponential potential.
Section 4 contains a full stability analysis of all critical
points in generalized assisted inflation. In section 5 the model
is generalized by the inclusion of nonzero spatial curvature and
finally in section 6 we summarize and conclude. In the appendices
we prove some results.

\section{The model}\label{system}

The scalars are denoted by $\phi_i$ and are assembled in an
$N$-vector $\phi$. The model is defined by the following action
\begin{equation}
S=\int d^4x\,\sqrt{-g}\Bigl[\frac{1}{2\kappa^2
}\mathcal{R}-\tfrac{1}{2}\langle\partial_{\mu}\phi\,,\,\partial^{\mu}\phi\rangle-V(\phi)
\Bigr]+ S_M \, ,
\end{equation}
where $\kappa^2=8\pi G$ with $G$ Newton's constant and
$\langle\partial_{\mu}\phi\,,\,\partial^{\mu}\phi\rangle$ is
shorthand for
$\sum_{i=1}^N\partial_{\mu}\phi_i\partial^{\mu}\phi_{i}$. The
potential $V(\phi)$ is a sum of $M$ exponential terms
\begin{equation}
V(\phi)=\sum_{a=1}^M\Lambda_a\,\text{exp}[-\kappa\,\langle\alpha_a\,,\,\phi\rangle],
\end{equation}
where
$\langle\alpha_a\,,\,\phi\rangle=\sum_{i=1}^{N}\alpha_{ai}\phi_i$.
There are $M$ vectors $\alpha_a$ with $N$ components
$\alpha_{ai}$. The indices $i,j,\ldots$ run from 1 to $N$ and
denote the components of the vectors $\phi$ and $\alpha_a$. The
indices $a, b,\ldots$ run from 1 to $M$ and label the different
vectors $\alpha_a$ and the constants $\Lambda_a$. $S_M$ is the
action for a barotropic fluid with density $\rho$ and pressure $P$
and as equation of state $P=(\gamma-1)\rho$. It is assumed that
the barotropic fluid respects the strong energy condition, which
means that $2/3<\gamma<2$.

In this paper we make use of linear field redefinitions. If the
scalars transform linearly as $\phi \longrightarrow\phi'=S\phi$,
where $S$ is an element of GL$(\mathbb{R},N)$ then the vectors
$\alpha_a$ transform in the dual representation
$\alpha_a\longrightarrow \alpha_a'=S^{-T}\alpha_a$. This can be
seen from the definition of $\alpha_a'$
\begin{equation}
\langle \alpha_a\,,\,\phi\rangle\equiv\langle
\alpha_a'\,,\,\phi'\rangle\,.
\end{equation}
Field redefinitions that shift the scalar fields leave the
$\alpha_a$ invariant, but change the $\Lambda_a$.

From the action we can deduce some properties of this system by
looking at transformations in scalar space. The kinetic term is
invariant under constant shifts  and $O(N)$-rotations of the
scalars. These transformations map the multiple exponential
potential to another multiple exponential potential but with
different $\Lambda_a$ and $\alpha_a$. Such redefinitions do not
alter the physics, they just rewrite the equations. Therefore
qualitative features only depend on $O(N)$-invariant combinations
of the $\alpha_a$-vectors (for example $\langle\alpha_a\,,\,
\alpha_b \rangle$). By shifting the scalars we can always rescale
$R$ of the $\Lambda_a$ to be $\pm 1$, where $R$ is the number of
independent $\alpha$-vectors.

The Ansatz for the metric is that of a flat FLRW-universe and
accordingly the scalars can only depend on the cosmic time $\tau$
\begin{equation}
 ds^2=-d\tau^2+a(\tau)^2d{\vec x}^{\,2}\, ,\quad \phi_i=\phi_i(\tau)\,.
\end{equation}
In section \ref{curvature} the analysis is extended to spatially
curved FLRW-universes. The equations of motion are
\begin{align}
&H^2=\tfrac{\kappa^2}{3}\Bigl[\tfrac{1}{2}\langle\dot{\phi}\,,\,\dot{\phi}\rangle+V(\phi)+\rho \Bigr]\, ,\label{Friedmann}\\
&\ddot{\phi}_i+3H\dot{\phi}_i+\partial_i V(\phi)=0\, ,\\
&\dot{\rho}+3\gamma H\rho=0\, ,\label{continu}
\end{align}
where $H=\dot{a}/a$ is the Hubble constant. The dot denotes
differentiation with respect to $\tau$. The equations
(\ref{Friedmann}-\ref{continu}) are referred to as the Friedmann
equation, the Klein--Gordon equation and the continuity equation,
respectively.

One can rewrite equations (\ref{Friedmann}-\ref{continu}) as an
autonomous system. We define the following dimensionless variables
\begin{equation}
X_i=\frac{\kappa\dot{\phi}_i}{\sqrt{6}\,H},\quad
Y_a=\frac{\kappa^2}{{3\,H^2}}\,
\Lambda_a\,\text{exp}[-\kappa\,\langle\alpha_a\,,\,\phi\rangle],\quad
\Omega=\frac{\kappa^2\rho}{3H^2}\, .
\end{equation}
If we write $X^2=\sum_{i}X_i^2$ and $Y=\sum_aY_a$ the equations of
motion become
\begin{align}
&\Omega+X^2+Y-1=0\label{FRIEDMANNII}\, ,\\
&X_i'=3X_i\Bigl(-1+X^2+\tfrac{\gamma}{2}\Omega\Bigr)+\sqrt{\tfrac{3}{2}}\sum_{a}\alpha_{ai}Y_a\, ,\label{Xprime}\\
&Y'_a=Y_a\Bigl(
-\sqrt{6}\,\,\langle\alpha_a\,,\,X\rangle+6X^2+3\gamma\Omega\Bigr)\,
,\label{Yprime}\\
&\Omega'=3\Omega\Bigl(2X^2+\gamma(\Omega-1)\Bigr)\label{OMEGAPRIME}\,,
\end{align}
where the prime denotes differentiation with respect to $\ln(a)$
\footnote{The use of $\ln(a)$ as a time coordinate fails when
$\dot{a}=0$. This is no problem for studying critical points and
their stability because in the neighborhood around a critical
point there always exists a region where the coordinate is well
defined. In reference \cite{Heard:2002dr} an explicit example is
given where $\dot a$ becomes zero at some point.}.

It can be shown that if $\Omega+X^2+Y-1=0$ initially then
equations (\ref{Xprime}-\ref{OMEGAPRIME}) guarantee that it is
satisfied at all times.  Hence, given the correct initial
conditions the dynamics is described by equations
(\ref{Xprime}-\ref{OMEGAPRIME}).

The number of linearly independent vectors $\alpha_a$ is denoted
by $R$. If $R<N$ one can rotate the scalars such that
$\phi_{R+1},\ldots,\phi_N$ no longer appear in the potential
($\alpha_{ai}=0$ for $i>R$). These scalars are then said to be
decoupled or free.

In the next section we will construct all critical point solutions
of the autonomous system (\ref{Xprime}-\ref{OMEGAPRIME}). For that
purpose it is useful to consider the matrix $A$
\begin{equation}
A_{ab}=\langle\alpha_a\,,\,\alpha_b\rangle\,.
\end{equation}
The models are divided into two classes. The first class is
defined by $R=M$ and the second class by $R<M$. Algebraically the
two differ in the following way
\begin{align}
& 1.\;\;R=M \quad \Longleftrightarrow \quad\text{det}A > 0 \,,\\
& 2.\;\;R<M \quad \Longleftrightarrow \quad\text{det}A = 0.
\end{align}
The first possibility, where $A$ is invertible, is called
generalized assisted inflation.

Assisted inflation is the subclass where $A$ is diagonal. This
implies that in assisted inflation the $\alpha_a$ are
perpendicular to each other and that one can choose an orthonormal
basis in which $\alpha_{ai}=\alpha_{a}\,\delta_{ai}$. In that
basis the potential becomes
\begin{equation}\label{assisted}
V(\phi)=\sum_{a=1}^M\Lambda_a\,\text{exp}[-\kappa\,\alpha_a\,\phi_a]\,.
\end{equation}
It is this particular form of the potential that is referred to as
assisted inflation in the literature. We want to emphasize that
the latter definition is basis-dependent. Potentials different
from (\ref{assisted}) but with a diagonal matrix $A$ can be
brought to the form (\ref{assisted}) through an $O(N)$-rotation of
the scalar fields. An $O(N)$-invariant definition of assisted
inflation is that $A$ is a diagonal matrix.

In any system with multiple fields but a single exponential such
as studied in \cite{vandenHoogen:2000cf}, the matrix $A$ is
trivially diagonal. One can perform a rotation on the scalars such
that only one scalar appears in the potential and all the others
are decoupled. In order to have a system whose scalars are
mutually interacting one needs at least two exponential terms both
containing more than one scalar.

\section{The critical points}\label{criticalpoints}

Critical points are defined as solutions of the autonomous system
for which $\Omega'=X_i'=Y_a'=0$. From the acceleration equation,
\begin{equation}
\dot{H}=-\tfrac{\kappa^2}{2}\Bigl[\langle\dot{\phi}\,,\,\dot{\phi}\rangle+\gamma\rho
\Bigr]\,,
\end{equation}
it follows that in a critical point $\dot{H}/H^2$ is constant. If
the constant differs from zero we put $\dot{H}/H^2\equiv -1/p\,$
and then the scale factor becomes
\begin{equation}
    a(\tau)=a_{0}\,\big(\frac{\tau}{\tau_{0}}\big)^{p}\,.
\end{equation}
In terms of the dimensionless variables $p$ can be expressed as
\begin{equation}\label{power}
p=\frac{1}{3(X^2+\tfrac{\gamma}{2}\Omega)}\,.
\end{equation}
When $\dot{H}/H^2=0$ the scale factor is
\begin{equation}
a(\tau)=a_0\,e^{H\tau}\,,
\end{equation}
and space-time is de Sitter.

The requirement that $Y'_a=0$ can be satisfied in two ways as can
be seen from
\begin{equation}
Y'_a=Y_a\Bigl(
-\sqrt{6}\,\,\langle\alpha_a\,,\,X\rangle+6X^2+3\gamma\Omega\Bigr)\,.
\end{equation}
Either $Y_a=0$ or the second factor on the right hand side equals
zero. If we put $Y_a=0$ by hand and then solve for the $X_i$ and
the remaining $Y_a$, the critical point is called a
\emph{nonproper critical point}. If we put the second factor to
zero by hand and then solve for $X_i$ and $Y_a$, the critical
point is called a \emph{proper critical point}. The name nonproper
is given since a critical point with some $Y_a=0$ has
$\infty$-valued scalar fields. Therefore these critical points are
no proper solutions to the equations of motion, they are
asymptotic descriptions of solutions. The proper critical points
generically have nonzero $Y_a$ and therefore are proper solutions
to the equations of motion. But in some cases one finds that
$Y_a=0$ for proper critical points, although one did not put those
$Y_a$ to zero by hand.

Regardless of whether critical points are proper solutions to the
equations of motion, they are all equally important in providing
information about the orbits. That is, they are either repellers,
attractors or saddle points.

\subsection{Generalized assisted inflation}

\subsubsection*{Proper critical points}

First we construct the proper critical points by proceeding as
explained in \cite{Collinucci:2004iw} and find
\begin{equation}\label{CRITICAL1}
Y_a=2\frac{3p-1}{3p^2}\sum_{b}[A^{-1}]_{ab}\,, \quad
X_i=\frac{1}{p}\sqrt{\tfrac{2}{3}}\sum_{ab}\alpha_{ai}[A^{-1}]_{ab}\,,\quad
\Omega=1-\frac{2}{p}\sum_{ab}[A^{-1}]_{ab}\,.
\end{equation}
The value of $p\,$ is found by combining equations
(\ref{FRIEDMANNII},\,\ref{power},\,\ref{CRITICAL1}). One gets
\begin{equation}\label{POWER2}
p^2-p\,(2\sum_{ab}[A^{-1}]_{ab}+\tfrac{2}{3\gamma})+\tfrac{4}{3\gamma}\sum_{ab}[A^{-1}]_{ab}=0\,.
\end{equation}
The two possible solutions are
\begin{equation} \label{power3}
p_{\phi}=2\sum_{ab}[A^{-1}]_{ab}\quad \text{and} \quad
p_{\phi+\rho}=\frac{2}{3\gamma}\,.
\end{equation}
The first solution has vanishing barotropic fluid ($\Omega=0$) and
is called \emph{the scalar-dominated solution}. The second
solution is a \emph{matter-scaling
solution}\cite{Copeland:1997et}; the ratio of the fluid density
$\Omega$ to the scalar field density
$\Omega_{\phi}\equiv\kappa^{2}\rho_{\phi}/3H^{2}$ (with
$\rho_{\phi}\equiv\tfrac{1}{2}\langle\dot{\phi}\,,\,\dot{\phi}\rangle+V$)
is finite and constant. For the matter-scaling solution the scalar
field mimics the fluid resulting in $p=\tfrac{2}{3\gamma}\,$,
which is the same powerlaw for a universe filled with only the
barotropic fluid.

In terms of the scalar fields the solutions read
\begin{equation}\label{expliciet}
\phi_i=\frac{\sqrt{6}X_i\,p}{\kappa}\ln|\tau|+c_i\,.
\end{equation}

\subsubsection*{Nonproper critical points}

To construct nonproper critical points we put a certain subset of
$\,Y_{\bar{a}}\,$ equal to zero by hand and then solve for the
$X_i$ and the remaining $Y_a$. We denote the elements of the
subset of $Y_a$ that were put to zero with barred indices
$\bar{a}$, so $Y_{\bar{a}}=0$.

The solutions are still given by equations (\ref{CRITICAL1}) and
(\ref{power3}) but now the indices $a,b,c$ run over the subset of
nonzero $Y_a$ and the matrix $A$ has to be replaced by the matrix
$A(\bar{a},\bar{b},\ldots)$, which is the submatrix of $A$ formed
by deleting the $\bar{a},\bar{b},\ldots$ columns and rows of $A$.

Similar to the proper critical points, the nonproper critical
points are divided into matter-scaling solutions and
scalar-dominated solutions. If all $Y_a=0$ the Friedmann equation
(\ref{FRIEDMANNII}) gives that $X^2+\Omega=1$. From
(\ref{OMEGAPRIME}) we notice that either $\Omega=0$ or $X^2=0$.
For the first case where $\Omega=0$, the solution is called a
\emph{kinetic-dominated solution} and $p=1/3$. For the second case
where $\Omega=1$, the solution corresponds to a universe only
filled with a barotropic fluid and is termed the
\emph{fluid-dominated solution} ($p_{\rho}=2/3\gamma$).

The maximal number of critical points can be seen to be
\begin{equation}
\text{Max}\, \sharp \, \text{critical points} = 2\times 2^M.
\end{equation}
The reason is that there are $2^M$ different ways of putting $Y_a$
equal to zero.  For each combination there are two possibilities:
when not all $Y_a$ equal zero there is the scalar-dominated and
the matter-scaling solution and when all $Y_a=0$ there is the
fluid-dominated solution and the kinetic-dominated solution.

For the scalar-dominated solutions the nonproper critical points
correspond to universes with an expansion that is always slower
than that of the proper critical point. This is due to the
following lemma.

Consider an invertible matrix $A$ with as entries innerproducts of
vectors $A_{ab}=\langle\alpha_a\,,\, \alpha_b\rangle$. Define
$A(\bar{a},\bar{b},\ldots)$ as before, then we have the following
inequality
\begin{equation}\label{lemma}
\sum_{ab}[A^{-1}]_{ab}\geq\sum_{cd}[A(\bar{a},\bar{b},\ldots)^{-1}]_{cd}\,.
\end{equation}
For diagonal $A$ this property is obvious. The proof for general
$A$ can be found in appendix \ref{proof}.

\subsubsection*{Existency conditions}

The critical points constructed above do not always exist. For
instance, it is clear from the definition of the $Y_a$-variables
that they must have the same sign as the $\Lambda_a$. So, if the
solution for $Y_a$ does not respect that, it has to be
discarded\footnote{In \cite{Bergshoeff:2003vb} it is shown that
critical points that violate the existency conditions can still
play a rol in understanding the late time behavior of general
solutions.}. Of all critical points two always exist, the
kinetic-dominated and fluid-dominated solution. For the other
critical points it depends on whether they are matter-scaling or
scalar-dominated solutions.

For the existence of a matter-scaling solution there is an extra
condition coming from $\Omega\geq 0$. For the matter-scaling
solution the conditions $\Omega\geq 0$ and
$\text{sgn}(Y_a)=\text{sgn}(\Lambda_a)$ become (\ref{CRITICAL1})
\begin{equation}\label{existence}
\sum_{ab}[A^{-1}]_{ab} \leq \frac{1}{3\gamma}\,,\quad
\sum_{b}A^{-1}_{ab}\gtrless 0 \quad \text{for}\quad \Lambda_a
\gtrless 0\, .
\end{equation}
This shows for instance that for a single negative exponential
potential there is no matter-scaling solution \cite{Heard:2002dr}.

If the critical point is a scalar-dominated solution then
existence is guaranteed if
\begin{equation}\label{existence2}
(\frac{1}{p_{\phi}}-3)\sum_b [A^{-1}]_{ab}\gtrless 0 \quad
\text{for}\quad \Lambda_a \gtrless 0\,.
\end{equation}
For the nonproper critical points the same equations hold by
changing $A$ to $A(\bar{a},\bar{b}\ldots)$. \\

To conclude, in generalized assisted inflation we distinguish four
kinds of critical points: the (non)proper scalar-dominated
solutions, the (non)proper matter-scaling solutions, the
fluid-dominated solution and the kinetic-dominated solution. The
kinetic and fluid-dominated solutions always exist and for the
matter-scaling and the scalar-dominated solutions the existency
conditions are given in (\ref{existence}) and (\ref{existence2}),
respectively.

\subsection{The $R<M$ case}\label{RsmallerthanM}

This section is brief since details can be found in
\cite{Collinucci:2004iw}.

An important difference with the previous case where $R=M$ is that
these potentials can have stationary points $\partial_i V(\phi)=0$
which correspond to de Sitter solutions ($a\sim e^{\tau}$) when
$V>0$. Since $R<M$ the $\alpha_a$-vectors are linearly dependent
and we can always choose an independent set of $R$ vectors
$\alpha_a$ with $a=1\ldots R$. The remaining vectors $\alpha_b$
($b=R\ldots M$) can be expressed as linear combinations of that
independent set
\begin{equation}
\alpha_b=\sum_{a=1}^R c_{ba}\alpha_a\,.
\end{equation}
Concerning the proper critical points one has to distinguish
between the case of potentials with \emph{affine}
coupling\footnote{In \cite{Collinucci:2004iw} we called this
coupling a convex coupling. Later we found that the name affine is
more appropriate.} and those without. By affine coupling we mean
the situation where we can find a set of independent $\alpha_a$'s
such that the coefficients $c_{ba}$ obey
\begin{equation}
\sum_a c_{ba}=1 \,\,\, \text{for all}\,\,\,b\,\,.
\end{equation}
These couplings are for instance found in group manifold
reductions of pure gravity \cite{Bergshoeff:2003vb}. If the
couplings are affine then the proper critical points correspond to
powerlaw solutions.

Since the inverse of the matrix $A$ does not exist when $R<M$ the
formulas for the solutions differ. We introduce the symmetric
matrix $B$
\begin{equation}
B_{ij}=\sum_{a=1}^M\alpha_{ai}\alpha_{aj}\,.
\end{equation}
If we assume that a basis rotation is performed whereby the
decoupled scalars are removed from the potentials we can restrict
the index $i$ to run from 1 to $R$. The decoupled scalars obey
$X_R=\ldots=X_N=0$. With this restriction on the index $i$ the
matrix $B$ becomes an $R\times R$ invertible matrix. The powerlaw
solutions when the coupling is affine is given by
\begin{equation}
X_i=\frac{\sqrt{2/3}}{p}\,\sum_{ja} B^{-1}_{ij}\alpha_{aj}\,.
\end{equation}
For $p$ there are again two possibilities
\begin{equation}
p_{\phi}=2\sum_i\Bigl(\sum_{ja}[B^{-1}]_{ij}\alpha_{aj}\Bigr)^2\quad
\text{and} \quad p_{\phi+\rho}=\tfrac{2}{3\gamma}\,.
\end{equation}
Therefore there exist proper scaling solutions (matter-scaling and
scalar-dominated) when the number of exponential terms exceeds the
number of scalars, but only if the coupling is affine. When $R<M$
and the coupling is not affine there are no proper scaling
solutions, but there can still exist nonproper scaling solutions
(which are referred to as approximate scaling solutions in
reference \cite{Karthauser:2006wb}).

When the couplings are not affine the only possible proper
critical point is a de Sitter universe.

For the construction of the nonproper critical points a subset of
the $Y_a$ must equal zero. Depending on the subset that is put to
zero, the resulting equations are that of the $R=M$ or $R<M$-case.
So in general nonproper critical points in the $R<M$ case can be
de Sitter and powerlaw.

The existence of the critical points is more complicated than for
generalized assisted inflation since the $Y_a$-variables are no
longer independent variables but are nonlinear dependent.

\section{Stability analysis}\label{stability}

To determine the stability of a critical point one linearizes the
equations of motion around the critical point. The matrix that
appears in the linearized autonomous equations is called the
stability matrix. Stability of the critical point requires that
the real parts of the eigenvalues of the stability matrix are
negative.

We will not discuss the stability of critical points for models
with $R<M$ but only for $R=M$. The reason is the difficulty that
arises from the dependence of the $Y_a$. It is a hard problem to
write down the independent perturbations around a critical point.
There exist exceptions where the perturbations are independent
when $R=M-1$, see \cite{Li:2005ay} for an example where
$R=1,\,M=2$.

Assume $(X_i,Y_a,\Omega)$ is a critical point and consider a
perturbation $(X_i+\delta_i,Y_a+\delta_a,\Omega+\delta_{\Omega})$.
If we substitute the perturbations in equations (\ref{Xprime}) and
(\ref{Yprime}) and keep the linear terms we find
\begin{align}
&\delta_i'=(\tfrac{1}{p}-3)\delta_i+3(2-\gamma)\sum_j X_iX_j\delta_j-\sum_a(\tfrac{3}{2}\gamma X_i\delta_a -\sqrt{\tfrac{3}{2}}\alpha_{ai}\delta_a)\,,\label{stability1}\\
&\delta_a'=Y_a\sum_j
\Bigl(6(2-\gamma)X_j-\sqrt{6}\alpha_{aj}\Bigr)\delta_j+
\delta_a\Bigl(-\sqrt{6}\langle\alpha_a\,,\,X \rangle +
\tfrac{2}{p}-3\gamma Y_a\Bigr)\,.\label{stability2}
\end{align}
The $\delta_{\Omega}$ perturbations are obtained by varying the
Friedmann equation (\ref{FRIEDMANNII}). Under this assumption it
is not an independent perturbation. The equations above can be
written as a matrix equation $\delta'=M\,\delta$. From
(\ref{stability1},\,\ref{stability2}) one can read off the values
of the stability matrix $M$
\begin{align}
& M_{ij}=[\tfrac{1}{p}-3]\mathbbm{1}_{ij}+3(2-\gamma)X_iX_j\,,\nonumber \\
& M_{ib}=-\tfrac{3\gamma}{2}X_i+\sqrt{\tfrac{3}{2}}\alpha_{bi}\,, \\
& M_{aj}=6(2-\gamma)Y_aX_j-\sqrt{6}\alpha_{aj}Y_a\,, \nonumber\\
& M_{ab}=\mathbbm{1}_{ab}\Bigl( -\sqrt{6} \langle\alpha_a \,,\,
X\rangle + \tfrac{2}{p} \Bigr) - 3\gamma Y_a\nonumber\,.
\end{align}

We now discuss the decoupled scalars $\phi_{R+1},\ldots,\phi_N$.
Since $\alpha_{aR+1}=\ldots=\alpha_{aN}=0$ it follows from
equation (\ref{Xprime}) that either $X_{R+1}=\ldots=X_N=0$ or some
$X_{i>R}\neq 0$. The latter only happens for the kinetic-dominated
solution ($X^2=1,\Omega=0$). The effect of free scalars at the
kinetic-dominated solution is discussed in
\cite{Bergshoeff:2003vb}. For all critical points except the
kinetic-dominated solution $M$ has only diagonal components in the
directions $i>R$:
\begin{equation}
M_{ib}=M_{aj}=0\,\,,\,M_{ij}=(\tfrac{1}{p}-3)\delta_{ij}\quad\text{for}\quad
i>R\,.
\end{equation}
The decoupled scalars decouple in the stability-analysis and do
not introduce instabilities as long as $p>1/3$.

We first investigate the eigenvalues of the stability matrix at
the nonproper critical points and then at the proper critical
points.

\subsection{Stability of nonproper critical points}\label{nonproper}

The submatrix $M_{ij}$ is diagonalizable with an
$O(R)$-transformation that rotates the scalars. The eigenvalues
are
\begin{equation}
E_1=[\tfrac{1}{p}-3]\quad,\quad
E_2=[\tfrac{1}{p}-3]+3(2-\gamma)X^2\,.
\end{equation}
The multiplicity of $E_1$ is $R-1$ and the multiplicity of $E_2$
is $1$. This is sufficient to understand the stability of the
critical points for which all $Y_a=0$; the kinetic-dominated and
the fluid-dominated solution. For these two solutions we then
diagonalize the $M_{ij}$-submatrix  via an $O(R)$-rotation and we
end up with a matrix $M$ which is upper triangular. The
eigenvalues can then be read off from the diagonal. For the
fluid-dominated solution the eigenvalues are $3(\gamma/2-1)<0$ and
$3\gamma>0$. Therefore the fluid-dominated solution is always
unstable. For the kinetic-dominated solution the eigenvalues are
$3(2-\gamma)>0$ and $ -\sqrt{6} \langle\alpha_a \,,\, X\rangle +
2/p$, the first eigenvalue is always positive and hence the
kinetic-dominated solution is always unstable.

Assume a certain nonproper critical point has one $Y_a$ equal to
zero and all other $Y_a$ nonzero. After renumbering we can choose
it to be $Y_1$. The $Y_1$-perturbation decouples from all the
other perturbations. The reason is that if $Y_1=0$ the $a=1$ row
of $M$ only contains its diagonal element $M_{a=1\,a=1}$ and
therefore the determinant of $M-\lambda$ becomes
\begin{equation}
\text{det}(M-\lambda)=(M_{a=1\,a=1}-\lambda)\text{det}(\tilde{M}-\lambda)\,,
\end{equation}
where $\tilde{M}$ denotes the matrix obtained after deleting the
$a=1$ row and $a=1$ column from $M$. The matrix $\tilde{M}$ is the
same as the stability matrix of the proper critical point
belonging to the new system that one obtains by removing the
$\Lambda_1\,\text{exp}[-\kappa\langle\alpha_1\,,\,\phi\rangle]$-term
from the potential.

This means that the stability of the nonproper critical point with
$Y_1=0$ and $Y_{a>1}\neq 0$ boils down to checking the sign of
$M_{a=1\,a=1}$ and to study the stability of the proper critical
point of the truncated system $\tilde{M}$. The stability of proper
critical points is left for the next section and we now focus on
the sign of $M_{a=1\,a=1}$.

Stability requires
\begin{equation}
M_{a=1\,a=1}=-\sqrt{6}\langle\alpha_1\,,\,X\rangle+2/p <0\,.
\label{stabiliteit}
\end{equation}
In appendix \ref{proof} we prove that this expression is
equivalent to
\begin{equation}\label{ong1}
\sum_b[A^{-1}]_{1b}<0\,.
\end{equation}

For nonproper critical points with more $Y_a$'s equal to zero,
$Y_1=Y_2=\ldots=Y_C=0$ we find $C$ conditions
\begin{equation}\label{ong2}
\sum_{b=1}^{R-C+1}[A(2,\ldots,C)]^{-1}_{1b}<0\,\,\,,\ldots
\,,\sum_{b=1}^{R-C+1}[A(1,\ldots,C-1)]^{-1}_{1b}<0\,.
\end{equation}

If $\sum_b[A^{-1}]_{ab}>0$  for all $a=1\ldots M$ then equation
(\ref{ong2}) cannot be satisfied (see appendix \ref{proof}) and
there does not exist a stable truncation of the $Y_a$. In the case
of a diagonal $A$ matrix this property is obvious, hence we find
that in assisted inflation nonproper critical points are always
unstable.

\subsection{Stability of proper critical points}\label{proper}

\subsubsection*{The characteristic polynomial}

If we consider perturbations around a proper critical point a
simplification occurs in the stability matrix. The $M_{ab}$
submatrix reduces to $M_{ab}=-3\gamma Y_a$, since now
\begin{equation}\label{innerproduct}
\langle\alpha_a\,,\,X\rangle=\tfrac{2}{\sqrt{6}p}\, .
\end{equation}

The eigenvalues of $M$ are found by solving the characteristic
polynomial det$(M-\lambda \mathbbm{1})=0$. The computation is
simplified using an LU-decomposition of $M-\lambda \mathbbm{1}$.
The characteristic polynomial becomes
\begin{equation}
\text{det}(M_{ij}-\lambda\mathbbm{1}_{ij})\,\text{det}(Z_{ab})=0\,,
\end{equation}
where we defined the matrix $Z$
\begin{equation}
Z_{ab}=M_{ab}-\lambda\mathbbm{1}_{ab}-M_a^{\,\,i}[(M-\lambda\mathbbm{1})^{-1}]_{i}^{\,\,j}M_{jb}\,.
\end{equation}

The LU-decomposition is not valid when $\lambda$ is an eigenvalue
of $M_{ij}$ ($E_{1,2}$). Solutions to det$(Z)=0$ are eigenvalues
of $M$ if they differ from $E_{1,2}$. With this LU-decomposition
one has to check separately whether $E_{1,2}$ are eigenvalues of
$M$.

The matrix $Z_{ab}$ contains only contracted $i$-indices and is
therefore $O(R)$-invariant. We can choose a specific orthonormal
basis to compute it. The easiest choice is that basis in which
$M_{ij}$ is diagonal, where we have
\begin{equation}
M_a^{\,\,i}[(M-\lambda\mathbbm{1})^{-1}]_i^{\,\,j}M_{jb}=\frac{1}{E_1-\lambda}M_a^{\,\,i}M_{ib}+(\frac{1}{E_2-\lambda}-\frac{1}{E_1-\lambda})M_{aR}M_{Rb}\,.
\end{equation}
This expression contains $X_R$ and $\alpha_{aR}$ in the specific
basis. One can determine them explicitly since by definition of
the new basis $X=(0,\ldots,0,\sqrt{X^2})$. Using this and the fact
that expression (\ref{innerproduct}) is $O(R)$-invariant we find
that $\alpha_{aR}=2/p\sqrt{6X^2}$ and hence
\begin{equation}\label{Xmatrix}
Z_{ab}=cY_a+\frac{3}{E_1-\lambda}A_{ab}Y_a-\lambda\delta_{ab}\,,
\end{equation}
with $c$ the following expression\footnote{For the special case of
$R=1$ the formulae can be taken over if one replaces $E_1$ by
$E_2$ everywhere.}
\begin{equation}
c= -3\gamma + \frac{1}{E_2-\lambda}\Bigl[ 9(2-\gamma)\gamma
X^2+\frac{3\gamma-12}{p}+\frac{2}{p^2X^2}\Bigr]
-\frac{2}{p^2X^2(E_1-\lambda)} \,.
\end{equation}
If one adds all rows to the first row and then subtracts the first
column from all the other columns the new matrix $Z'$ one obtains
has the same determinant as $Z$ and is given by
\begin{equation}
Z'=\left[\begin{array}{cc}
cY+\frac{6p-2}{p^2}\frac{1}{E_1-\lambda}-\lambda &  0 \\
cY_a+\frac{3}{E_1-\lambda}A_{a1}  &
\frac{3}{E_1-\lambda}(A_{ab>1}-A_{a1})-\delta_{ab}\lambda
\end{array}\right]\,.
\end{equation}\\
This shows that the determinant factorizes in a $c$-dependent and
a $c$-independent polynomial
\begin{equation}\label{decompositionforc}
\text{det}(Z')=\Bigl(cY+\frac{6p-2}{p^2}\frac{1}{E_1-\lambda}-\lambda\Bigr)
\text{det}\Bigl(\frac{3}{E_1-\lambda}(A_{ab}-A_{a1})-\delta_{ab}\lambda\Bigr)
\,\, \text{with}\,\, a,b>1\,.
\end{equation}

The factorization of the characteristic polynomial in a
$c$-dependent and a $c$-independent part allows us to divide the
perturbations into fluid perturbations and scalar field
perturbations. The fluid perturbations are those that give rise to
the eigenvalues that come from the $c$-dependent factor whereas
the scalar field perturbations belong to those eigenvalues that
come from the $c$-independent factor.

\subsubsection*{Fluid perturbations}

If we put the first factor of (\ref{decompositionforc}) to zero we
get the following eigenvalues for the scalar-dominated solution
\begin{equation}\label{eigenwaardenscalardominated}
\lambda_1=E_1 \, , \quad \lambda_2=\frac{2}{p}-3\gamma\,,
\end{equation}
and for the matter-scaling solution
\begin{equation}
\lambda_{\pm}=\lambda=\frac{3}{4}(2-\gamma)\Bigl[-1\pm\sqrt{1-8\gamma
\frac{1-3\gamma\sum_{ab}[A^{-1}]_{ab}}{2-\gamma}}\Bigr]\,.
\end{equation}

These eigenvalues have a simple interpretation. Stability of the
scalar-dominated solution requires
\begin{equation}
\lambda_2<0 \quad \Longleftrightarrow \quad p>2/3\gamma\quad
\Longleftrightarrow \quad
p_{\phi}>p_{\phi+\rho}.\label{eigenwaarde12}
\end{equation}
For stability it does not matter whether $E_1$ is an eigenvalues
or not, because its implication for stability ($p>1/3$) is less
strong then the implications coming from $\lambda_2$
($p>2/3\gamma$) when we assume that $2/3<\gamma<2$.

Stability of the matter-scaling solution requires
\begin{equation}
\text{Re}\lambda_{\pm}<0 \quad \Longleftrightarrow \quad
p>2\sum_{ab}[A^{-1}]_{ab}
 \quad\Longleftrightarrow \quad p_{\phi+\rho}>p_{\phi}.\label{eigenwaarde+-}
\end{equation}
The existency condition for the matter-scaling solution
(\ref{existence}) implies stability (\ref{eigenwaarde+-}) which in
turn implies instability of the scalar-dominated solution. In
\cite{Tsujikawa:2006mw} it is proven that for lagrangians that
allow for scaling solutions but without cross-coupling, the
scalar-dominated solution is unstable when the matter-scaling
solution is stable.

\subsubsection*{Scalar field perturbations}

The remaining eigenvalues follow from the smaller polynomial
\begin{equation}\label{moeilijk}
\text{det}\Bigl(\frac{3}{E_1-\lambda}(A_{ab}-A_{a1})-\delta_{ab}\lambda\Bigr)=0
\quad \text{with}\quad a,b>1\,.
\end{equation}
Determining the solutions of (\ref{moeilijk}) for arbitrary $A$ is
a hard problem and we will use another technique to determine
stability. This technique does not use the (linearized) autonomous
equations and was developed in the context of assisted inflation
\cite{Liddle:1998jc} and elaborated upon in \cite{Malik:1998gy}.
Below we will extend this to the case of generalized assisted
inflation.

The technique is based upon making a basis rotation in scalar
space such that one direction corresponds to the critical point
solution and the other directions are constructed perpendicular to
it. Then the idea is to check whether the critical point direction
is a minimum of the potential with respect to the other direction,
if so, the critical point is an attractor.

In order to understand which basis rotation to make notice that
the proper critical points are given by
\begin{equation}
\phi_i(\tau)=\frac{\sqrt{6}X_ip}{\kappa}\ln|\tau|+c_i\,,
\end{equation}
irrespective of whether it is a matter-scaling or a
scalar-dominated solution. Combinations of the form
\begin{equation}
\frac{\phi_i}{X_i}-\frac{\phi_j}{X_j}\,,
\end{equation}
are constant in time at the critical point. Therefore we define
the following new scalar combinations
\begin{align}
&\phi'_i=N_i[\frac{\phi_i}{X_i}-\frac{\phi_{i+1}}{X_{i+1}}] \qquad \text{for} \quad i<R\,,\\
&\phi'_R=N_R[\frac{\phi_1}{X_2X_3\ldots
X_R}+\frac{\phi_2}{X_3X_4\ldots
X_RX_1}+\ldots+\frac{\phi_R}{X_1X_2\ldots
X_{R-1}}]\,,\label{phi''}
\end{align}
with $N_i$ and $N_R$ normalization constants given by
\begin{equation}
 N_i= \Bigl[X_i^{-2}+X_{i+1}^{-2}\Bigr]^{-1/2}\,,\quad
 N_R=X^{-1}\,\Pi_{i=1}^R\,X_i\,.
\end{equation}

The directions $\phi'_{i<R}$ are all perpendicular to the
direction $\phi'_R$, but they are not perpendicular among
themselves, hence this is not an orthogonal field redefinition. To
establish orthogonality we perform a Gram--Schmidt procedure on
the $\phi'_{i<R}$ directions. The Gram--Schmidt procedure takes
linear combinations of the $\phi'_{i<R}$ to obtain new directions
$\phi''_{i<R}$ that are mutually perpendicular. It follows from
the construction that the $\phi''_{i<R}$ are also constant at the
critical point. We define $\phi''_R\equiv \phi'_R$.

The transition of the $\phi_i$ to the $\phi''_i$ is done
orthogonally and therefore the system still has a canonical
kinetic term and a multiple exponential potential
\begin{equation}
S=\int d^4x\,\sqrt{-g}\Bigl[\frac{1}{2\kappa^2
}\mathcal{R}-\tfrac{1}{2}\langle\partial_{\mu}\phi''\,,\,\partial^{\mu}\phi''\rangle-\sum_{a=1}^M\Lambda_a\,\text{exp}[-\kappa\,\langle\alpha''_a\,,\,\phi''\rangle]\,
\Bigr] \,.
\end{equation}

We focus on the appearance of $\phi''_R$ in the potential, so we
have to find what $\alpha''_{aR}$ is. The scalars are redefined by
an orthogonal transformation, $\phi''=O\phi$. The $\alpha$-vectors
transform as dual vectors $\alpha''_a=[O^{-1}]^T\,\alpha_a$ which
simplifies to $\alpha''_a=O\,\alpha_a$ for orthogonal
transformations. In components this means that
\begin{equation}
\alpha''_{ai}=O_i^{\,\,j}\alpha_{aj}\,.
\end{equation}
To find $\alpha''_{aR}$ it then suffices to construct
$O_R^{\,\,\,i}$ for all $i$.

Although we do not know the matrix $O$ since we did not specify
the Gram--Schmidt procedure, we can find $O_R^{\,\,i}$ from
(\ref{phi''}) since $\phi'_R=\phi''_R$
\begin{equation}
O_R^{\,\,i}=\frac{X_i}{X}\,.
\end{equation}
Thus in the potential the scalar $\phi''_R$ appears in the $a$-th
exponent as
\begin{equation}
\alpha''_{aR}\phi''_R\,=\,X^{-1}\langle
X\,,\,\alpha_a\rangle\,\,\phi''_R\,=\,\frac{2\phi''_R}{\sqrt{6}\,p\,X}\,.
\end{equation}
The product $\alpha''_{aR}\phi''_R$ is independent of $a$, so all
exponentials have the same coupling to $\phi''_R$ and we can
rewrite the potential as
\begin{equation}\label{new potential}
V(\phi'')=e^{\frac{2\phi''_R}{\sqrt{6}\,p\,X}}\,W(\phi''_1,\ldots,\phi''_{R-1})\,,\quad
W(\phi''_1,\ldots,\phi''_{R-1})=\sum_{a=1}^M\Lambda_a\,e^{-\kappa\,\langle\alpha''_a\,,\,\phi''\rangle}\,,
\end{equation}
and $\langle\alpha''_a\,,\,\phi''\rangle$ does not contain
$\phi''_R$;
\begin{equation}
\langle\alpha''_a\,,\,\phi''\rangle=\sum_{i=1}^{R-1}\alpha''_{ai}\phi''_{i}\,.
\end{equation}

The fact that we were able to rewrite the potential like (\ref{new
potential}) extends the results of \cite{Malik:1998gy} to the case
of generalized assisted inflation.

The function $W(\phi''_1,\ldots,\phi''_{R-1})$ in (\ref{new
potential}) contains $R-1$ scalar fields and has $R$ exponential
terms. These functions can have a stationary point ($\partial
V=0$) and this stationary point is unique. Moreover this
stationary point corresponds to the critical point, the solution
for which the fields $\phi''_1,\ldots,\phi''_{R-1}$ take constant
values. This is proven by considering the Klein--Gordon equations
in the new basis
\begin{equation}
\ddot{\phi''}_i+3H\dot{\phi''}_i+\partial_i
(e^{\frac{2\phi''_R}{\sqrt{6}\,p\,X}}\,\,W) =0\,.
\end{equation}
Since for the critical point $\phi''_1,\ldots,\phi''_{R-1}$ are
constant we see that
\begin{equation}
\partial_{i<R}(W(\phi''_1,\ldots,\phi''_{R-1}))=0\,.
\end{equation}

The critical point is an attractor if and only if this stationary
point is a minimum of $W$. Stability requires that the matrix
$\partial_i\partial_j W(\phi''_1,\ldots,\phi''_{R-1})$ (for
$i,j=1\ldots,R-1$) is positive definite. If all $\Lambda_a>0$ then
$\partial_i\partial_j W$ is positive definite since
$\partial_i\partial_j W=\alpha^T\eta\alpha$ with $\eta$ a diagonal
matrix with positive entries,
$\eta_{aa}=\Lambda_a\text{exp}[-\kappa\langle\alpha''_a\,,\,\phi''\rangle]>0$.
Hence if the proper critical point of a potential with all
$\Lambda_a>0$ exists, it is stable. When some $\Lambda_a<0$ we
have proven in appendix \ref{proof2} that the matrix
$\,\partial_i\partial_j W=\alpha^T\eta\alpha$ is not positive
definite if $p>1/3$. Hence potentials with some negative
$\Lambda_a$ cannot have stable \emph{proper} critical points with
$p>1/3$.

\subsection{Summary of the stability analysis}

We discuss qualitatively the results first for assisted inflation
and then for generalized assisted inflation. We explain how to
find the stable critical points (attractors) by eliminating all
unstable critical points.

\subsubsection*{Assisted inflation}

At the end of section \ref{nonproper} we found that in assisted
inflation all nonproper critical points are unstable. So if an
attractor is present it is either the proper matter-scaling or the
proper scalar-dominated solution.

Because of the existency conditions for critical points with
diagonal $A$ (\ref{existence},\,\ref{existence2}), the proper
critical point only exist when all $\Lambda_a>0$\footnote{ We
assume that $p>1/3$, since scalar-dominated solutions with
$p_{\phi}>1/3$ are unstable because one can proof that $E_1=1/p-3$
is always an eigenvalue of the stability matrix for assisted
inflation.}. This implies that for assisted inflation models with
some negative $\Lambda_a$ there is no attractor, all critical
points are unstable.

In subsection \ref{proper} we analyzed the eigenvalues coming from
the $c$-dependent factor of the characteristic polynomial
det$(Z)=0$. The stability conditions are $p_{\phi}>p_{\phi+\rho}$
for the scalar-dominated solution and $p_{\phi}<p_{\phi+\rho}$ for
the matter-scaling solution. This means that the stable proper
critical point is the one that represents the universe with the
fastest expansion of the two. The nonproper critical points are
all unstable and  have smaller $p$-values, which implies that the
attractor is the critical point with the fastest expansion of
\emph{all} the critical points. Furthermore,  when the
matter-scaling solution exists, it is stable and the
scalar-dominated solution is unstable.

In sum, assisted inflation models have an attractor when all
exponential terms are positive and furthermore the attractor is
unique and corresponds to the critical point with the fastest
expansion. Thus in assisted inflation the universe dynamically
favors a fast expansion!

A fast expansion is favored because of the `assisting behavior' of
the scalar fields. Each scalar field gives a positive contribution
to the Hubble friction and thus more scalar fields lead to more
friction and more friction leads to slow rolling fields. Therefore
one says that the scalars assist each other in helping the
universe to expand. The assisting behavior is proven for all
multi-field lagrangians that allow for scaling solutions
\emph{and} have no cross-coupling of the scalars
\cite{Tsujikawa:2006mw}.

\subsubsection*{Generalized assisted inflation}
We separately discuss the cases where all terms in the potential
are positive and where some are negative.

\begin{enumerate}
\item $\forall\,\,\Lambda_a>0$. If a proper critical point exists,
the scalar field perturbations are stable since all $\Lambda_a>0$.
As in assisted inflation the fluid perturbations imply that the
stable proper critical point is the one that represents the
universe with the fastest powerlaw, either the matter-scaling (if
$p_{\phi+\rho}>p_{\phi}$) or the scalar-dominated solution (if
$p_{\phi}>p_{\rho+\phi}$). Since the proper critical point exists
we have $\sum_b[A^{-1}]_{ab}>0$ for all $b$\,\footnote{We use that
$p_{\phi}>1/3$ for scalar-dominated solutions of positive
potentials.}, which implies that truncations are unstable (see
discussion below (\ref{ong2})). Hence the nonproper critical
points are unstable and the attractor is unique. The cases
analyzed in \cite{Green:1999vv} are of this type.

If there exists no proper critical point, attractors have to be
nonproper. To find stable nonproper critical points, the stability
of the truncation (\ref{ong1},\ref{ong2}) is the only condition
that has to be fulfilled. Therefore nonproper critical points can
be stable when the proper critical point does not exist
\footnote{But the kinetic- and fluid-dominated solution can never
be stable as we showed in section 4.1.}. It is not guaranteed that
there is a unique stable truncation nor if a stable truncation is
possible at all. Hence the number of attractors is case dependent
and can be zero, one or higher.

\item $\exists \,\,\Lambda_a<0$\,.  Whereas critical points with
negative $Y_a$ do not exist in assisted inflation, they can exist
in generalized assisted inflation but are unstable as we now
explain. If $Y_{n_1},\ldots, Y_{n_q}$ are negative at a critical
point $\wp$ than so are $\Lambda_{n_1},\ldots,\Lambda_{n_q}$ in
order for $\wp$ to exist. If $\wp$ is proper we proved at the end
of section 4.2 that it must be unstable. If $\wp$ is nonproper
i.e. $Y_{s_1}=\ldots=Y_{s_p}=0$ then we showed in section 4.1 that
the $Y_{s_1}\,,\ldots,Y_{s_p}$-perturbations decouple from all
other perturbations. The latter perturbations are the same
perturbations as that of a proper critical point $\tilde{\wp}$ of
a lagrangian where $\Lambda_{s_1}\,,\ldots,\Lambda_{s_p}$ are put
to zero. This truncated lagrangian contains the negative
$\Lambda_{n_1},\ldots,\Lambda_{n_q}$ hence $\tilde{\wp}$ is
unstable. Since the perturbations around $\wp$ contain the
perturbations around $\tilde{\wp}$, $\wp$ must also be unstable.
This proves that proper and nonproper critical points with
negative $Y_a$ are unstable.

Assume $\Lambda_{1},\ldots,\Lambda_{n}<0$ and all other
$\Lambda_a>0$. If some of the $Y_{1},\ldots,Y_{n}$ differ from
zero they are negative in order for the critical point to exist.
Therefore we have to truncate at least all $Y_{1},\ldots,Y_{n}$ to
find a stable critical point. If such truncations are unstable,
there is no attractor. Again, it is not guaranteed that there is a
unique stable truncation nor if a stable truncation is possible at
all.
\end{enumerate}

\section{The effects of spatial curvature}\label{curvature}

We show that for critical points with $k=0$ curvature
perturbations do not affect the eigenvalues found in section 4.
Instead curvature perturbations lead to an additional eigenvalue.

The metric Ansatz is
\begin{equation}
ds^{2}=-d\tau^{2}+a(\tau)^{2}\big(\frac{dr^{2}}{1-kr^{2}}+r^{2}d\Omega^{2}\big),
\end{equation}
where $d\Omega^{2}$ is the metric on the unit 2-sphere and
$k=0,\pm 1$ is the normalized curvature. The equations of motion
are
\begin{align}
&H^2=\tfrac{\kappa^2}{3}\Bigl[\tfrac{1}{2}\langle\dot{\phi}\,,\,\dot{\phi}\rangle+V(\phi)+\rho \Bigr]-\frac{k}{a^{2}}\, ,\label{kFriedmann}\\
&\ddot{\phi}_i+3H\dot{\phi}_i+\partial_i V(\phi)=0\, ,\label{kKleinGordon}\\
&\dot{\rho}+3\gamma H\rho=0\, .\label{kcontinuity}
\end{align}

Using the same dimensionless variables as in section \ref{system},
the equations of motion become
\begin{align}
&\Omega+X^2+Y-1=\frac{k}{{\dot a}^{2}}\label{kFRIEDMANNII}\, ,\\
&X_i'=X_i(Q-2)+\sqrt{\tfrac{3}{2}}\sum_{a}\alpha_{ai}Y_a\, ,\label{kXprime}\\
&Y'_a=Y_a\Bigl(
-\sqrt{6}\,\,\langle\alpha_a\,,\,X\rangle+2(Q+1)\Bigr)\,
,\label{kYprime}\\
&\Omega'=\Omega\Bigl(2Q-(3\gamma-2))\Bigr)\label{kOMEGAPRIME}\,.
\end{align}
where in the last line a new quantity $Q$, the deceleration
parameter, is introduced. It is defined by
\begin{equation}
Q\equiv-\frac{a\ddot a}{{\dot
a}^2}=2X^{2}-Y+\Omega(\frac{3\gamma}{2}-1)\,.
\end{equation}
For powerlaw solutions $a\sim\tau^{p}$ the deceleration parameter
is simply $Q=\frac{1-p}{p}$.

One might wonder if the system is still autonomous due to the
$\frac{k}{{\dot a}^{2}}$-term. Using equations
(\ref{kXprime}-\ref{kOMEGAPRIME}) one shows that
\begin{equation}
    \big(X^{2}+Y+\Omega-1\big)'=2Q\big(X^{2}+Y+\Omega-1\big)\,.
\end{equation}
The solution to this equation is
\begin{equation}
    \vert X^{2}+Y+\Omega-1\vert=\frac{C}{{\dot a}^{2}},
\end{equation}
with $C\geq 0$. If equation (\ref{kFRIEDMANNII}) is satisfied at
the instant $\tau=\tau_{0}$, it is satisfied at all times. We
conclude that the evolution never changes the value of $k$ and
that the dynamics is still described by the autonomous system
(\ref{kXprime}-\ref{kOMEGAPRIME}).

The critical points can be divided into critical points with $k=0$
and critical points with $k\neq 0$. The critical points with $k=0$
are the same as the ones discussed in section
\ref{criticalpoints}. The critical points with $k\neq 0$
necessarily have $p=1$.

\subsection*{Curvature perturbations}

We define the function $f(X,Y,\Omega)$ by
\begin{equation}
    f(X,Y,\Omega)=X^2+Y+\Omega-1.
\end{equation}
The function $f$ provides us with information about the spatial
curvature, for example $\mbox{sgn}f=k$. Consider perturbations
$X_i+\delta_i$, $Y_a+\delta_a$ and $\Omega+\delta_{\Omega}$. They
induce a perturbation $\delta f$ which we will call a curvature
perturbation. It satisfies the following differential equation
\begin{equation}\label{curvatureperturb}
    (\delta f)'=2Q\delta f+2f\delta Q.
\end{equation}
In section \ref{stability} ($f=0$) we considered perturbations
that satisfy $\delta f=0$. In order to study the stability of the
critical point against curvature perturbations we need to treat
$\delta_i$, $\delta_a$ and $\delta_{\Omega}$ as independent
perturbations. The stability matrix $M$ can be calculated in
exactly the same manner as in section \ref{proper} with the only
difference that the resulting matrix has one extra row and one
extra column. The entries are
\begin{equation}
\begin{array}{l@{\quad\quad\quad}l}
    M_{\Omega\Omega}=2Q+(3\gamma-2)(\Omega-1)\,,& M_{\Omega j}=8\Omega X_{j}\,,\\
    M_{a\Omega}=(3\gamma-2)Y_{a}\,, & M_{i\Omega}=(\tfrac{3}{2}\gamma-1)X_{i}\,,\\
    M_{aj}=-\sqrt{6}Y_{a}\alpha_{aj}+8Y_{a}X_{j}\,,& M_{\Omega b}=-2\Omega\,,\\
    M_{ij}=(Q-2)\mathbbm{1}_{ij}+4X_{i}X_{j}\,,& M_{ib}=-X_{i}+\sqrt{\tfrac{3}{2}}\alpha_{bi}\,,\\
    M_{ab}=-2Y_{a}+\big(2(Q+1)-\sqrt{6}\sum_{i}\alpha_{ai}X_{i}\big)\mathbbm{1}_{ab}\,.&\label{entriesofM}
\end{array}
\end{equation}
Performing an LU-decomposition of the matrix
$M-\lambda\mathbbm{1}$ one shows that
\begin{equation}\label{LUdecomposition}
    \mbox{det}(M-\lambda\mathbbm{1})=(M_{\Omega\Omega}-\lambda)\mbox{det}(M_{ij}-\lambda\mathbbm{1}_{ij})\mbox{det}(C_{ij})
    \mbox{det}(Z_{ab})=0,
\end{equation}
where $C_{ij}$ and $Z_{ab}$ are given by
\begin{align}
    C_{ij}&=\mathbbm{1}_{ij}-\sum_{k}(M_{ik}-\lambda\mathbbm{1}_{ik})^{-1}\frac{M_{k\Omega}M_{\Omega
    j}}{M_{\Omega\Omega}-\lambda}\,,\\
    Z_{ab}&=M_{ab}-\lambda\mathbbm{1}_{ab}-\frac{M_{a\Omega}M_{\Omega
    b}}{M_{\Omega\Omega}-\lambda} -\sum_{l,k,n}C_{nk}^{-1}
    (M_{kl}-\lambda\mathbbm{1}_{kl})^{-1}\big(M_{an} - \frac{M_{a\Omega}M_{\Omega n}}{M_{\Omega\Omega}-\lambda}\big)\big(M_{lb} - \frac{M_{l\Omega}M_{\Omega
    b}}{M_{\Omega\Omega}-\lambda}\big)\nonumber.
\end{align}
This decomposition is valid if
\begin{equation}
    \lambda\neq M_{\Omega\Omega}\,,\ \ \lambda\neq E_{1}\,,\ \ \lambda\neq
    E_{2}\,, \ \ \mbox{det}(C_{ij})\neq 0\,,
    \label{lambdacondities}
\end{equation}
where $E_{1}=Q-2$, $E_{2}=Q-2+4X^{2}$ are the eigenvalues of
$M_{ij}$. Making use of the basis in which
$X=(0,\ldots,0,\sqrt{X^{2}})$ as explained in section
\ref{stability} we find
\begin{align}
    \mbox{det}(C_{ij}) & =1-4\Omega\frac{3\gamma-2}{E_{2}-\lambda}\frac{
    X^2}{M_{\Omega\Omega}-\lambda}\,, \\
    Z_{ab} & =cY_{a}+\frac{3}{E_{1}-\lambda}A_{ab}Y_{a}-\lambda\mathbbm{1}_{ab}\,,
\end{align}
where the constant $c$ is
\begin{align}
    c=&-2+2\Omega\frac{3\gamma-2}{M_{\Omega\Omega}-\lambda}-\frac{2}{E_{1}-\lambda}\frac{(Q+1)^{2}}{X^{2}}-
    \frac{1}{\mbox{det}C_{ij}}\frac{2}{E_{2}-\lambda}\Bigl[5(Q+1) - 4X^{2}\nonumber\\
    &-\frac{(Q+1)^{2}}{X^{2}}
    -5(Q+1)\Omega\frac{3\gamma-2}{M_{\Omega\Omega}-\lambda}+8\Omega X^{2}\frac{3\gamma-2}{M_{\Omega\Omega}-\lambda} -
    4X^{2}\Omega^{2}\left(\frac{3\gamma-2}{M_{\Omega\Omega}-\lambda}\right)^{2}\Bigr]\,.
\end{align}

The matrix $Z$ is of the same form as in section \ref{proper} but
with a different constant $c$. In section \ref{proper} it was
shown that for proper critical points the determinant of $Z_{ab}$
factorizes in a $c$-dependent and a $c$-independent polynomial.
This result also applies to proper critical points with $k\neq 0$.
Therefore the only eigenvalues that change are those that are
related to $c$. These eigenvalues are determined by
\begin{equation}\label{cdependenteigenvalues}
    \lambda=cY+\frac{6p-2}{p^{2}}\frac{1}{E_{1}-\lambda}\,.
\end{equation}

This proves that the effects of curvature only manifest themselves
in a subset of the eigenvalues of the stability matrix. In this
sense the effects of curvature decouple from the scalar dynamics.

Next we discuss the effect of curvature perturbations for the
scalar-dominated and scaling solutions with $k=0$ as well as the
stability of the new critical points with $k\neq 0$. In both cases
we only consider proper critical points. For the case of nonproper
critical points the arguments of section \ref{nonproper} apply
here as well.

\subsection*{Stability of critical points with $k=0$}

For the scalar-dominated solution  we find the eigenvalue
\begin{equation}
    \lambda_{1}=M_{\Omega\Omega}=\frac{2}{p}-3\gamma,
\end{equation}
directly from the expression for $M$ (\ref{entriesofM}). Equation
(\ref{cdependenteigenvalues}) is solved by
\begin{equation}
    \lambda_{2}=E_{1}\,, \quad \lambda_{3}=\frac{2}{p}-2.
\end{equation}
The value of $\lambda_2$ violates the conditions
(\ref{lambdacondities}), however $\lambda_3 < 0$ requires $p>1$,
which is a stronger constraint than $E_1<0$.

For the matter-scaling solution equation
(\ref{cdependenteigenvalues}) gives
\begin{equation}
    \lambda_{1,2}=\frac{3}{4}(2-\gamma)\Bigl[-1\pm\sqrt{1-8\gamma
\frac{1-3\gamma\sum_{ab}[A^{-1}]_{ab}}{2-\gamma}}\Bigr]\,,\quad
\lambda_{3}=3\gamma-2 \,,\quad \lambda_{4}=M_{\Omega\Omega} \,.
\end{equation}
Eigenvalue $\lambda_4$ violates the conditions
(\ref{lambdacondities}). However, the condition $\lambda_4<0$ is
equivalent to $\lambda_3<0$.

In both cases the eigenvalues $\lambda_{1,2}$ are also found in
section \ref{proper} and the eigenvalue $\lambda_{3}$ is
additional. The condition $\lambda_3<0$ implies
$Q<0\Leftrightarrow\ddot{a}>0$. That $Q<0$ has to be interpreted
as stability against curvature perturbations is clear from
equation (\ref{curvatureperturb}). It follows that the
inflationary critical points are stable against curvature
perturbations. The matter-scaling solution is never inflationary
since the fluid respects the strong energy condition.

\subsection*{Stability of critical points with $k\neq 0$}

For the proper critical points with $k\neq 0$ we have $Q=\Omega=0$
from which it follows $Y=2X^{2}=\frac{4}{3}\sum_{ab}A_{ab}^{-1}$.
The curvature can be found by computing $f(X,Y,\Omega) =
2\sum_{ab}A_{ab}^{-1}-1$.

The solutions to equation (\ref{cdependenteigenvalues}) are
\begin{equation}
    \lambda_{1,2}=-1\pm\sqrt{8\sum_{ab}A_{ab}^{-1}-3}.
\end{equation}
Since $\Omega=0$, $\lambda_{3}=2-3\gamma$ is an eigenvalue.
Stability requires that
\begin{equation}
    \gamma > \frac{2}{3}\,,\quad
    \sum_{ab}A_{ab}^{-1}<\frac{1}{2}.
\end{equation}
Hence the only stable fixed points are the ones that have $k=-1$.
These are the curvature-scaling solutions of reference
\cite{vandenHoogen:1999qq}.  By a curvature-scaling solution is
meant that the energy density that effectively describes the
effects of spatial curvature is a constant fraction of the total
energy density.

If one puts all $Y_{a}$ equal to zero, then all $X_{i}$ are zero
and the critical point is never an attractor and represents a
Milne patch of Minkowski space-time.

\section{Conclusion}

In this paper we study the stability of critical points in models
with multiple exponential potentials and multiple fields. The
critical points are separated in proper and nonproper critical
points. The proper critical points are exact solutions to the
equations of motion and the nonproper solutions are asymptotic
descriptions of solutions.  A stability analysis is carried out
for models that belong to the class of generalized assisted
inflation. In that class a critical point with $k=0$ is either a
kinetic-dominated solution, a fluid-dominated solution, a
matter-scaling solution or a scalar-dominated solution. The
critical points with $k\neq 0$ are either curvature-scaling
solutions or Milne universes.

For critical points with $k\neq 0$, the curvature-scaling solution
with $k=-1$ is the only stable solution.

For critical points with $k=0$, the stability analysis shows that
the fluid perturbations, curvature perturbations and scalar field
perturbations all decouple from each other.  The existence of a
matter-scaling solution implies that there exists a
scalar-dominated solution (when $p_{\phi}>1/3$). Only one of the
two can be stable against fluid perturbations and that is the one
with the fastest powerlaw. This implies that inflationary
scalar-dominated solutions are stable against fluid perturbations
since we assume that the barotropic fluid respects the strong
energy condition. Only the inflationary scalar-dominated solutions
are stable against curvature perturbations.

Proper critical points are stable against scalar field
perturbations in assisted inflation when all terms in the
potential are positive. If some terms in the potential are
negative there is no attractor. Since proper critical points have
a higher powerlaw than the nonproper solutions, the critical point
with the fastest expansion is the unique critical point that is
stable against scalar field and fluid perturbations. Furthermore,
if the critical point is inflationary it is stable against
curvature perturbations. Therefore in assisted inflation a fast
expansion is dynamically favorable.

For generalized assisted inflation the stability is more
complicated. There are three main differences with assisted
inflation. Firstly, nonproper critical points can be stable.
Secondly, potentials with negative $\Lambda_a$ can have stable
(nonproper) critical points, as long as the truncation of the
negative terms can be done in a stable manner. Thirdly, there can
exist unstable critical points with a higher powerlaw than the
stable critical points. In that case the fastest expansion is not
dynamically the most favorable.

Despite the name generalized assisted inflation our models are
more suited for present-day acceleration than early-universe
inflation. Multiple exponential potentials either give rise to
transient acceleration with too little $e$-foldings for early
universe inflation or there exists an inflationary attractor
without exit.

Regarding present-day acceleration the shortcomings of our model
are different. To avoid the cosmic coincidence problem, the period
where the universe enters a stage of acceleration and where the
energy density of the barotropic fluid is comparable to that of
the scalar fields has to be long enough. In our model such a
period arises when the universe moves from an unstable
matter-scaling solution towards the stable inflationary attractor
\cite{Barreiro:1999zs}. During that transition there is a period
with acceleration and comparable energy densities. But to make
that period long enough one has to tune initial conditions.
According to \cite{Kim:2005ne} this tuning becomes worse for
multiple fields. A way to avoid this problem is by allowing
interaction between the barotropic fluid and the scalar fields.
Then there arises the possibility of inflationary attractors with
nonzero energy density for the barotropic fluid
\cite{Billyard:2000bh}.

\section*{Acknowledgments}
We would like to thank Dieter Van den Bleeken for a useful
discussion. AP, TVR and DBW are supported by the European
Commission FP6 program MRTN-CT-2004-005104 in which AP, TVR and
DBW are associated to Utrecht University. The work of AP, TVR and
DBW is part of the research programme of the ``Stichting voor
Fundamenteel Onderzoek der Materie'' (FOM). JH is supported by the
``breedtestrategie'' grant of the University of Groningen.

\appendix

\section{Properties of the $A$-matrix }\label{proof}

Consider $R$ linearly independent vectors $\alpha_a$. Their dual
vectors $\beta^b$, that is,
$\langle\alpha_a\,,\,\beta^b\rangle=\delta_a^b$ are uniquely
determined when we demand that they all lie in
$\text{Span}\{\alpha_a,\,\,a=1\ldots R\}$. Then we have that
\begin{equation}\label{expr1}
    [A^{-1}]_{ab}=\langle\beta^a\,,\,\beta^b\rangle.
\end{equation}
The following $R-1$ vectors
\begin{equation}\label{betavgl}
\bar{\beta}^a=\beta^a-\frac{\langle\beta^R\
,\,\beta^a\rangle}{\langle\beta^R\,,\,\beta^R\rangle}\beta_R\quad\text{with}\quad
a=1,\ldots,R-1\,,
\end{equation}
lie in $\text{Span}\{\alpha_a,\,\,a=1\ldots R-1\}$ and are such
that
\begin{equation}\label{expr2}
[A(R)^{-1}]_{ab}=\langle\bar{\beta}^a\,,\,\bar{\beta}^b\rangle.
\end{equation}

\subsection*{Proof of the lemma}

The sum of all matrix elements of $A^{-1}$ is given by
\begin{equation}
\sum_{ab}[A^{-1}]_{ab}=\sum_{ab<R}\langle\beta^a\,,\,\beta
^b\rangle+2\sum_{a<R}\langle\beta^a\,,\,\beta^R\rangle+\langle\beta^R\,,\,
\beta^R\rangle\,.
\end{equation}
If we rewrite the $\beta$'s in terms of the $\bar{\beta}$'s via
(\ref{betavgl}) and use that $\langle\bar{\beta^a}\,,\,\beta^R
\rangle=0$, we find
\begin{align}
&\sum_{ab}[A^{-1}]_{ab}=\sum_{ab<R}\langle\bar{\beta^a}\,,
\,\bar{\beta^b}\rangle+ \frac{1}{\langle\beta^R\,,\,\beta^R
\rangle}\bigr\{\sum_{ab<R}\langle\beta^a\,,\,\beta^R\rangle\langle\beta^b\,,\,\beta^R
\rangle +
\nonumber\\
&+\sum_{a<R}\langle\beta^a\,,\,\beta^R\rangle\langle\beta^R\,,\,\beta^R
\rangle+\langle\beta^R\,,\,\beta^R\rangle\langle\beta^R\,,\,\beta^R\rangle\bigl\}\,.
\end{align}
This can be rewritten as
\begin{equation}
\sum_{ab}[A^{-1}]_{ab}=\sum_{ab<R}[A(1)^{-1}]_{ab}+\frac{
1}{\langle\beta^R\,,\,\beta^R\rangle}\Bigl[\sum_{a=1}^R\langle\beta^R\,,\,
\beta^a\rangle\Bigr]^2\,,
\end{equation}
which proves (\ref{lemma}) since the second term on the RHS is
positive.

\subsection*{Stability of truncations}

The condition for stability of truncating $Y_1$ is given by
(\ref{stabiliteit}). If we substitute
\begin{equation}
X_i=\frac{1}{p}\sqrt{\tfrac{2}{3}}\sum_{ab}\alpha_{ai}[A(1)^{-1}]_{ab}\,,
\end{equation}
in (\ref{stabiliteit}) the stability condition is given by
\begin{equation}\label{ong3}
\sum_{a,b>1}A_{1a}\langle\bar{\beta_a}\,,\,\bar{\beta_b}\rangle>1
\,.
\end{equation}
If we rewrite the $\bar{\beta}$ in term of the $\beta$ in
(\ref{ong3}) and then use the expressions
(\ref{expr1},\ref{expr2}) we get
\begin{equation}\label{ong3}
\sum_{a,b=1}^{M}A_{1a}\Bigl([A^{-1}]_{ab}-[A^{-1}]_{a1}[A^{-1}_{1b}]
\Bigr)>1 \,.
\end{equation}
From which (\ref{ong1}) follows.

Using (\ref{ong1}) one proofs the stability conditions
(\ref{ong2}) for truncating multiple $Y_a$ in an analogues way.

Using the expressions (\ref{expr1},\ref{expr2}) one can proof with
the same techniques as above that if
\begin{equation}
\sum_{a=1}^M[A^{-1}]_{ab}>0\,\,\,\text{for all b}\,,
\end{equation}
then (\ref{ong2}) can never be satisfied.

\section{Eigenvalues of $\partial_i\partial_j W$}\label{proof2}

Here we will proof that $\partial_i\partial_j W\mid_{\phi^{*}}$
has some negative values if some of the $\Lambda_a$ are negative.

We use $\phi^{*}$ to denote the value of the $M-1$ scalar fields
at the point where all first derivatives are zero,
$\partial_iW\mid_{\phi^{*}}=0$. For simplicity of notation we
define the numbers
$\eta_a=\Lambda_a\,\text{exp}[-\kappa\langle\alpha_a,\phi^{*}\rangle]\,\,$
and the $M\times M$-matrix $\eta$
\begin{equation}
\eta=\text{diag}(\eta_1,\eta_2,\ldots,\eta_M)\,.
\end{equation}
Similarly we define the vector $\vec{\eta}$ as the vector with $M$
components $\eta_a$.

In this notation vanishing of the first derivatives implies the
following $M-1$ equations
\begin{equation}\label{perpendicular}
 \vec{\alpha_{i}}\cdot \vec{\eta} =0\,,
\end{equation}
where we indicated the $M-1$ columns of the matrix $\alpha$ as
$\vec{\alpha}_{i}$. We further assume that the critical point we
have in mind has $p>1/3$ which implies that $\sum_a\eta_a>0$.

If the eigenvalues of the $(M-1)\times (M-1)$-matrix given by
\begin{equation}
\partial_i\partial_j W\mid_{\phi^{*}} = [\alpha^T\eta\alpha]_{ij} =
\sum_{a=1}^M \alpha_{ai}\alpha_{aj}\eta_a
\end{equation}
are positive $\phi^{*}$  is a stable minimum otherwise it is a
saddle or a maximum. If all $\eta_a>0$ it is straightforward to
check that all eigenvalues are positive.

Assume, after possible renumbering, that $\eta_1$ is negative,
then define the $M \times M$-matrix $\beta$ as the matrix $\alpha$
with an extra column at the end with all entries of that column
equal to $1$
\begin{equation}
\beta=\left(\begin{array}{cccc}
\alpha_{11}& \alpha_{12} & \ldots &1 \\
\alpha_{21}& \alpha_{22} & \ldots &1 \\
\ldots & \ldots & \ldots &1 \\
\alpha_{M1}& \alpha_{M2} & \ldots & 1
\end{array}\right)\,.
\end{equation}
The matrix $\alpha$ has maximal rank and therefore all columns are
linearly independent. The same holds for the matrix $\beta$. The
reason is that the columns of $\alpha$ are all perpendicular to
$\vec{\eta}$, whereas the last column with all components equal to
$1$ has the following innerproduct with $\vec{\eta}$
\begin{equation}
\langle(1,\ldots,1),\vec{\eta} \rangle=\sum_{i}\eta_i>0\,.
\end{equation}
Hence $(1,\ldots,1)$ must have a component in the
$\vec{\eta}$-direction and it cannot be a linear combination of
the columns of $\alpha$. This means that $\beta$ has maximal rank
and since it is square we can define its inverse $\beta^{-1}$.

If we calculate the entries of $M\times M$-matrix $\beta^T\eta
\beta$ using the identity (\ref{perpendicular}) we find that it
takes the form
\begin{equation}
\beta^T\eta\beta=\left(\begin{array}{cc}
 \alpha^T\eta\alpha & 0 \\
 0 & \sum_{a}\eta_a
\end{array}\right)\,.
\end{equation}
We notice that the spectrum of $\beta^T\eta\beta$ equals the
spectrum of $\alpha^T\eta\alpha$ plus the positive eigenvalue
$\lambda=\sum_{a}\eta_a$. If $\beta^T\eta\beta$ has a negative
eigenvalue so has $\alpha^T\eta\alpha$. Therefore it is enough to
prove that there exists a vector $\psi$ such that $\langle \psi
|\beta^T\eta\beta |\psi\rangle <0\,$. It is easily checked that
\begin{equation}
 \psi=\beta^{-1}\left(\begin{array}{c}
 1  \\
 0  \\
 \ldots \\
 0
\end{array}\right)
 \end{equation}
is such a vector.
\bibliography{citaties}
\bibliographystyle{utphysmodb}
\end{document}